\documentclass[10pt,twocolumn]{article}

\usepackage[bookmarks=false,bookmarksnumbered=true,hypertexnames=false,breaklinks=true,]{hyperref}
\hypersetup{
  pdfauthor={Bernardi, Kamps, Kiseleva, Mueller},
  pdftitle={The Continuous Cold Start Problem in e-Commerce Recommender Systems},
  pdfsubject={CoCoS, continuous cold start problem},
  pdfkeywords={CoCoS: Continuous cold start problem},
  pdfcreator={LaTeX with hyperref package},
  pdfproducer={pdflatex}
}
\usepackage{xcolor}
\hypersetup{
    colorlinks,
    linkcolor={red!50!black},
    citecolor={blue!50!black},
    urlcolor={black}
}

\usepackage{comment}
\usepackage{xspace}
\usepackage{graphicx}
\usepackage{multirow}
\usepackage{amsmath}
\usepackage{color}

\usepackage[square,comma,numbers,sort&compress,sectionbib]{natbib} 
\def\:{\hskip0pt} 
\usepackage{tabularx} 

\newcommand{\booking}{\href{http://www.booking.com}{Booking.com}}
\newcommand{\cocos}{CoCoS}

\newcommand{\tempheader}[1]{\textcolor{blue}{\textbf{#1}}}
\renewcommand{\tempheader}[1]{}

\title{\centering The Continuous Cold Start Problem\\ in e-Commerce Recommender Systems}

\author{\centering
\begin{tabular}{c}
Lucas Bernardi\(^1\), Jaap Kamps\(^2\), Julia Kiseleva\(^3\), Melanie J.I. Mueller\(^1\)\\[1ex]
\small\(^1\)Booking.com, Amsterdam, Netherlands. Email: \{lucas.bernardi, melanie.mueller\}@booking.com\\
\small\(^2\)University of Amsterdam, Amsterdam, Netherlands. Email: kamps@uva.nl\\
\small\(^3\)Eindhoven University of Technology, Eindhoven, Netherlands. Email: j.kiseleva@tue.nl\\[0.5ex]
\end{tabular}
}

\begin{document}

\maketitle
\begin{abstract}
Many e-commerce websites use recommender systems to recommend items to users.  When a user or item is new, the system may fail because not enough information is available on this user or item.  Various solutions to this `cold-start problem' have been proposed in the literature.  However, many real-life e-commerce applications suffer from an aggravated, recurring version of cold-start even for known users or items, since many users visit the website rarely, change their interests over time, or exhibit different personas.  This paper exposes the \textsl{Continuous Cold Start} (\cocos) problem and its consequences for content- and context-based recommendation from the viewpoint of typical e-commerce applications, illustrated with examples from a major travel recommendation website, Booking.com. 
\end{abstract}

\medskip\small\vspace{1mm}
\noindent
{\bf Terms:} \cocos: continuous cold start\newline
\normalsize

\section{Introduction\label{sec:intro}}

\begin{figure*}[!tb]
\centerline{%
\includegraphics[width=1.\textwidth]{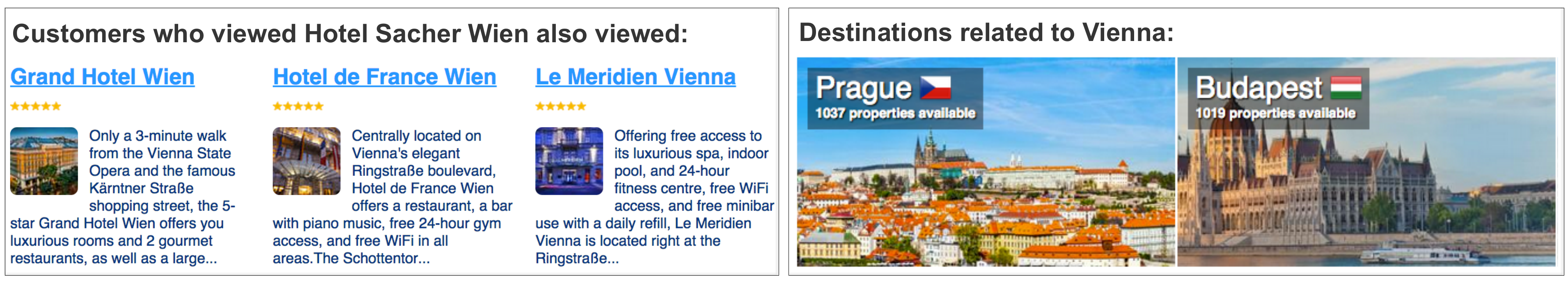}}
\caption{\label{fig:screenshots}Examples of recommender systems on \booking. User-to-user collaborative filtering (left): recommend accomodations viewed by similar users to a user who just looked at `Hotel Sacher Wien'. Item-to-item content-based recommendations (right): recommend destinations similar to a particular destination, Vienna. 
}
\end{figure*}

\tempheader{Intro recommender systems.}
Many e-\:commerce websites are built around serving personalized recommendations to users. Amazon.com recommends books, Booking.com recommends accommodations, Netflix recommends movies, Reddit recommends news stories, etc.  Two examples of recommendations of accomodations and destinations at \booking\ are shown in Figure~\ref{fig:screenshots}.
This widescale adoption of recommender systems online, and the challenges faced by industrial applications, have been a driving force in the development of recommender systems. The research area has been expanding since the first papers on collaborative filtering in the 1990s \cite{ResnickRiedl94, ShardanandMaes95}. Many different recommendation approaches have been developed since then, in particular content-based and hybrid approaches have supplemented the original collaborative filtering techniques \cite{AdomaviciusTuzhilin05}.

\tempheader{Classical cold-start problem.}
In the most basic formulation, the task of a recommender system is to predict ratings for items that have not been seen by the user. Using these predicted ratings, the system decides which new items to recommend to the user. Recommender systems base the prediction of unknown ratings on past or current information about the users and items, such as past user ratings, user profiles, item descriptions etc. If this information is not available for new users or items, the recommender system runs into the so-called \textit{cold-start problem}: It does not know what to recommend until the new, `cold', user or item has `warmed-up', i.e.\ until enough information has been generated to produce recommendations. For example, which accomodations should be recommended to someone who visits \booking\ for the first time? If the recommender system is based on which accomodations users have clicked on in the past, the first recommendations can only be made after the user has clicked on a couple of accomodations on the website. 

\tempheader{Patching the cold-start.}
Several approaches have been proposed and successfully applied to deal with the cold-start problem, such as utilizing baselines for cold users \cite{Kluver:2014:ERB:2645710.2645742}, combining collaborative filtering with content-based recommenders in hybrid systems \cite{schein2002methods}, eliciting ratings from new users \cite{RashidRiedls02}, or, more recently, exploiting the social network of users \cite{sedhain2014social, Guy:2009:PRS:1639714.1639725}. In particular, content-based approaches have been very successful in dealing with cold-start problems in collaborative filtering \cite{schein2002methods, BykauVelegrakis13, AharonSerfaty13, SaveskiMantrach14}.

These approaches deal explicitly with cold users or items, and provide a `fix' until enough information has been gathered to apply the core recommender system. Thus, rather than providing unified recommendations for cold and warm users, they temporarily bridge the period during which the user or item is `cold' until it is `warm'. This can be very successful in situations in which there are no warm users \cite{AharonSerfaty13}, or in situations when the warm-up period is short and warmed-up users or items stay warm. 

\tempheader{Getting into \cocos.}
However, in many practical e-\:commerce applications, users or items remain cold for a long time, and can even `cool down' again, leading to a \emph{continuous} cold-start (\cocos). In the example of \booking, many users visit and book infrequently since they go on holiday only once or twice a year, leading to a prolonged cold-start and extreme sparsity of collaborative filtering matrices, see Fig.~\ref{fig:users} (top). In addition, even warm long-term users can cool down as they change their needs over time, e.g.\ going from booking youth hostels for road trips to resorts for family vacations. Such cool-downs can happen more frequently and rapidly for users who book accommodations for different travel purposes, e.g.\ for leisure holidays and business trips as shown in Fig.~\ref{fig:users} (bottom). These continuous cold-start problems are rarely addressed in the literature despite their relevance in industrial applications. Classical approaches to the cold-start problem fail in the case of \cocos, since they assume that users warm up in a reasonable time and stay warm after that.  

\tempheader{Paper outline.}
In the remainder of the paper, we will elaborate on how \cocos\ appears in e-\:commerce websites (Sec.~\ref{sec:cocos}), outline some approaches to the \cocos\ problem (Sec.~\ref{sec:approaches}), and end with a discussion about possible future directions (Sec.~\ref{sec:discussion}).

\section{Continuous Cold-Start}\label{sec:cocos}

Cold-start problems can in principle arise on both the user side and the items side.

\subsection{User Continuous Cold-Start}

\begin{figure*}[!tb]
\centerline{%
\includegraphics[width=1.\textwidth]{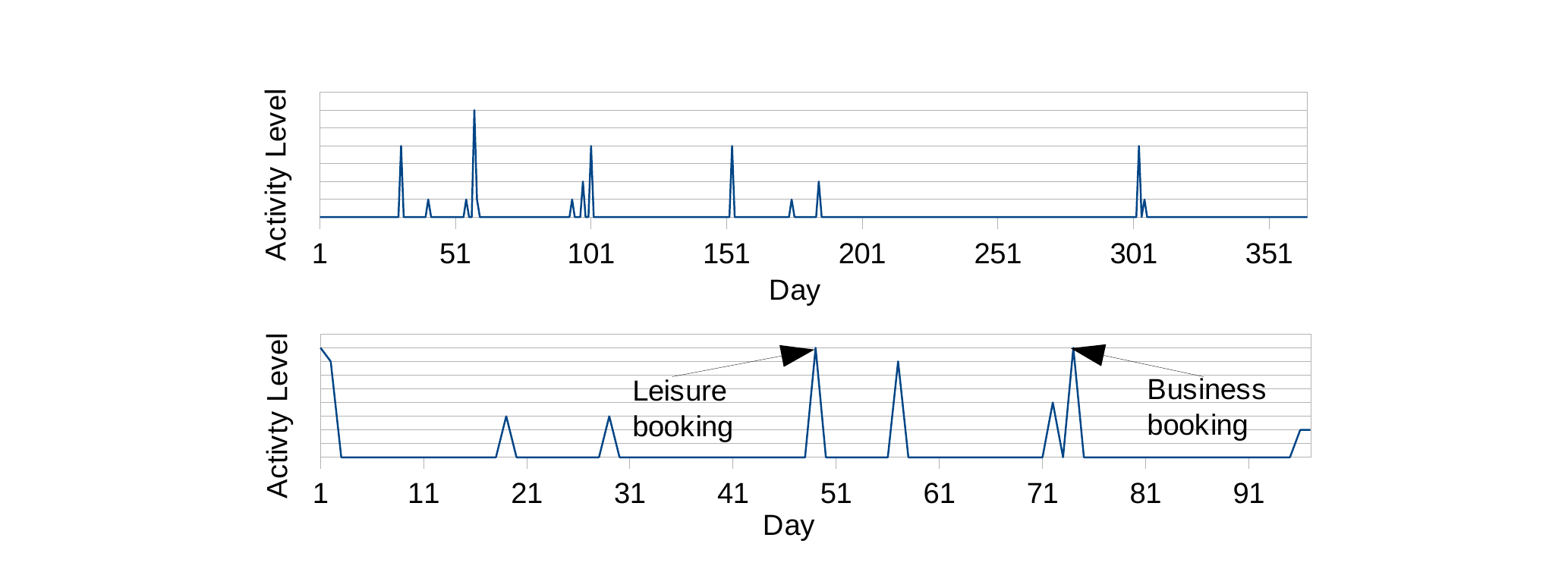}}
\caption{\label{fig:users}Continuously cold users at \booking. Activity levels of two randomly chosen users of \booking\ over time. The top user exhibits only rare activity throughout a year, and the bottom user has two different personas, making a leisure and a business booking, without much activity inbetween.
}
\end{figure*}

\tempheader{User \cocos.}
We first focus on the user side of \cocos, which can arise in the following cases:
\begin{description}
\item[Classical cold-start / sparsity:] new or rare users
\item[Volatility:] user interest changes over time
\item[Personas:] user has different interests at different, possibly close-by points in time
\item[Identity:] failure to match data from the same user
\end{description}
All cases arise commonly in e-\:commerce websites. New users arrive frequently (classical cold-start), or  may appear new when they don't log in or use a different device (failed identity match). Some websites are prone to very low levels of user activity when items are purchased only rarely, such as travel, cars etc., leading to sparsity problems for recommender systems. Most users change their interests over time (volatility), e.g.\ movie preferences evolve, or travel needs change. On even shorter timescales, users have different personas. Depending on their mood or their social context, they might be interested in watching different movies. Depending on the weather or their travel purpose, they may want to book different types of trips, see Figure~\ref{fig:users} for examples from \booking.

\begin{figure*}[!tb]
\centerline{%
\includegraphics[width=1.\textwidth]{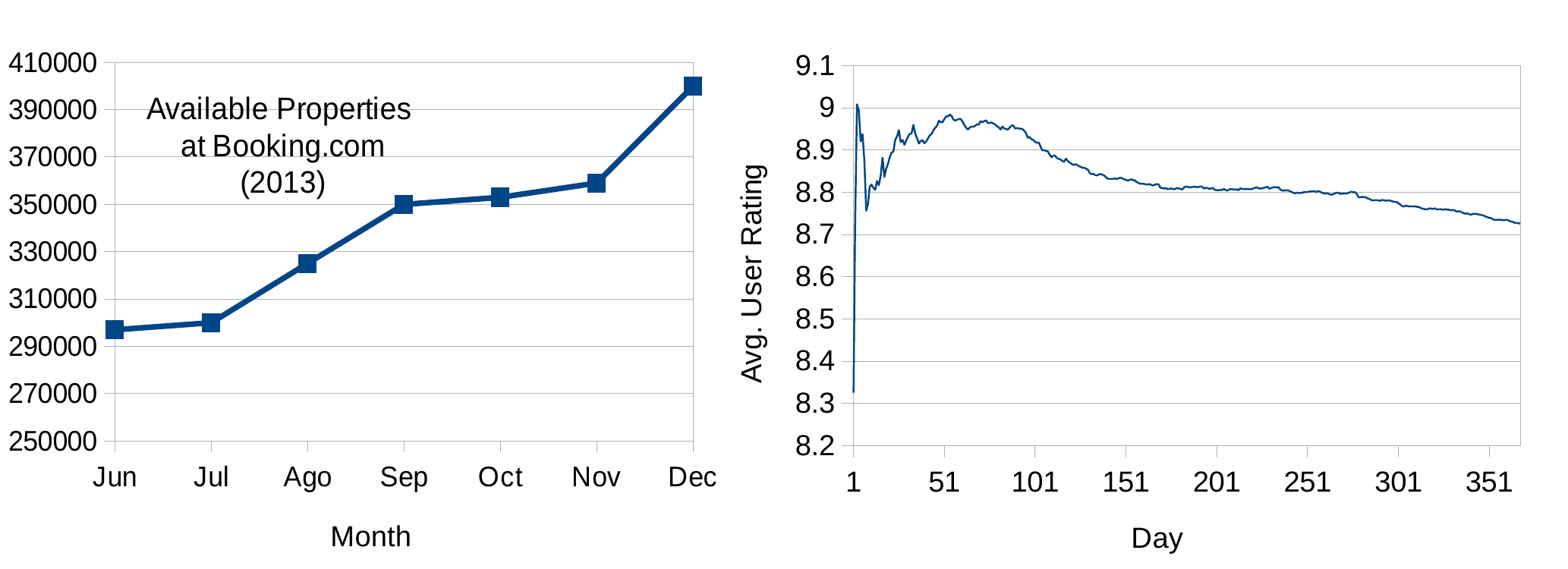}}
\caption{\label{fig:hotels}Continuously cold items at \booking. Thousands of new accommodations are added to \booking\ every month (left). The user ratings of a hotel can change continuously (right).
}
\end{figure*}

\tempheader{Collaborative vs.\ content-based.}
These issues arise for collaborative filtering as well as content-based or hybrid approaches, since both user ratings or activities as well user profiles might be missing, become outdated over time, or not be relevant to the current user persona.

\subsection{Item Continuous Cold-Start}

\tempheader{Symmetric \cocos\ for items.}
In a symmetric way, these \cocos\ problems also arise for items:
\begin{description}
\item[Classical cold-start / sparsity:] new or rare items
\item[Volatility:] item properties or value changes over time
\item[Personas:] item appeals to different types of users
\item[Identity:] failure to match data from the same item
\end{description}
New items appear frequently in e-\:commerce catalogues, as shown in Figure~\ref{fig:hotels} for accommodations at \booking. Some items are interesting only to niche audiences, or sold only rarely, for example books or movies on specialized topics. Items can be volatile if their properties change over time, such as s phone that becomes outdated once a newer model is released, or a hotel that undergoes a renovation. In the context of news or conversions, item volatility is also known as topic drift \cite{KnightsNicolov09}. Figure~\ref{fig:hotels} on the right shows fluctuations of the review score of a hotel at \booking. Some items have different `personas' in that they target several user groups, such as a hotel that caters to business as well as leisure travellers. When several sellers can add items to an e-\:commerce catalogue, or when several catalogues are combined, correctly matching items can be problematic (identity problem).

\section{Addressing Cold-Start}\label{sec:approaches}

Many approaches have been proposed to deal with the classical cold-start problem of new or rare users or items \cite{RashidRiedls02}. However, they mostly fail to address the more difficult \cocos. 

\tempheader{Why hybrid approaches fail.}
The most popular strategy to address the classic cold-start problem is the hybrid approach where collaborative filtering and content-based models are combined, see \cite{schein2002methods} as an example. If one of the two method fails due to a new user or item, the other method is used to `fill-in'. The most basic assumption is that similar users will like similar items. Similarity of users is measured by their purchase history when warm, and by their user profile when cold. Conversely,  similarities between items is computed by the set of users that purchased them when warm, and by their content when cold.
In \cocos, users change their interests, so both collaborative filtering and user-profile-based approaches can fail, since 
looking at the past and similarities can be misleading. Items also suffer from volatility, although to a lesser degree, which makes the standard hybrid approach also problematic for items.
Hybrid approaches also ignore the issue of multiple personas.

\tempheader{Previous work about \cocos}
Although, to our knowledge, the continuous cold-start problem as defined in this work has not been directly addressed in the literature, several approaches are promising. 

\citet{Tang:2014:ECB:2645710.2645732} propose a context-aware recommender system, implemented as a contextual multi-armed bandits problem. 
Although the authors report extensive offline evaluation (log based and simulation based) with acceptable CTR, no comparison is made from a cold-start problem standpoint.

\citet{sun2012dynamic} explicitly attack the user volatility problem. They propose a dynamic extension of matrix factorization where the user latent space is modeled by a state space model fitted by a Kalman filter. Generative data presenting user preference transitions is used for evaluation. Improvements of RMSE when compared to timeSVD \citep{koren2010collaborative} are reported. Consistent results are reported in \citep{chua2013modeling}, after offline evaluation using real data.
 
\citet{Tavakol:2014:FMD:2645710.2645739} propose a topic driven recommender system. At the user session level, the user intent is modeled as a topic distribution over all the possible item attributes. As the user interacts with the system, the user intent is predicted and recommendations are computed using the corresponding topic distribution. The topic prediction is solved by factored Markov decision processes. Evaluation on an e-\:commerce data set shows improvements when compared to collaborative filtering methods in terms of average rank.

\section{Discussion}\label{sec:discussion}

\tempheader{Summary.}
In this manuscript, we have described how \cocos, the continuous cold-start problem, is a common issue for e-\:commerce applications. Industrial recommender systems do not only have to deal with `cold' (new or rare) users and items, but also with known users or items that repeatedly `cool down'. Reasons for the recurring cool-downs include the volatility in user interests or item values, different personas depending on user context or item target audience, or identification problems due to logged-out users or items from different catalogues. Despite the practical relevance of \cocos, common literature approaches do not deal well with this issue. 

\tempheader{Suggested directions}
We consider several directions as particularly promising to deal with \cocos.  Traditional approaches to solve cold-start problems try to employ collaborative filtering based on pseudo or inferred clicks.
\tempheader{Social.}
Recommendations based on social networks are an interesting new development that can supplement missing information based on the social graph. For example, recommendations based on Facebook likes are proposed in \citep{sedhain2014social}. Beyond the difficulty to get access to social data, the application to user volatility or multiple personas remains challenging.
\tempheader{Online prediction.}
Online user intent prediction can be used to estimate a user's current profile on the fly. When a user visits the website, his browsing behavior is used to estimate his intent after a few clicks, which are then used to compute recommendations accordingly. However, this still delays recommendations until enough clicks have occurred, which can be problematic if quick recommendations are needed. For example, in last-minute bookings, users may be pressed to book an accommodation quickly, leading to very short sessions. 

More promising approaches employ content based or contextual recommendation.
\tempheader{Content.}
Content based recommendations can be very effective based on very little signal: just an initial query or single interaction can be exploited to find an initial item or set of items and exploit relations between items to make effective recommendations.
\tempheader{Context.}
In particular context aware recommendations are one of the most promising strategies when it comes to solving \cocos. In this setup, recommendations are computed based on the current context of the current visitor and the behaviour of other users in similar contexts \citep[see ][]{gediminasadomavicius2010, Shi_cikm_2014, Hawalah_2014} for examples. Context is defined as a set of features such as location, time, weather, device, etc. Often this data is readily available in most commercial implementations of recommender systems. This approach naturally addresses sparsity by clustering users into contexts. Since context is determined in a per-action basis, user volatility and multiple personas can be addressed robustly. 
On the other hand, context aware recommenders cannot address the item side of the problem and they might also suffer from cold-start problems in the case of a cold context that has never seen before by the system.


\footnotesize 
\bibliographystyle{abbrvnat}
\setlength{\bibhang}{1em}
\setlength{\bibsep}{0.5\itemsep} 

\begin{thebibliography}{20}
\providecommand{\natexlab}[1]{#1}
\providecommand{\url}[1]{\texttt{#1}}
\expandafter\ifx\csname urlstyle\endcsname\relax
  \providecommand{\doi}[1]{doi: #1}\else
  \providecommand{\doi}{doi: \begingroup \urlstyle{rm}\Url}\fi

\bibitem[Adomavicius and Tuzhilin(2005)]{AdomaviciusTuzhilin05}
G.~Adomavicius and A.~Tuzhilin.
\newblock Toward the next generation of recommender systems: a survey of the
  state-of-the-art and possible extensions.
\newblock \emph{IEEE Transactions on Knowledge and Data Engineering},
  17:\penalty0 734--749, 2005.

\bibitem[Adomavicius and Tuzhilin(2011)]{gediminasadomavicius2010}
G.~Adomavicius and A.~Tuzhilin.
\newblock Context-aware recommender systems.
\newblock In \emph{Recommender Systems Handbook}, pages 217--253, 2011.

\bibitem[Aharon et~al.(2013)Aharon, Aizenberg, Bortnikov, Lempel, Adadi,
  Benyamini, Levin, Roth, and Serfaty]{AharonSerfaty13}
M.~Aharon, N.~Aizenberg, E.~Bortnikov, R.~Lempel, R.~Adadi, T.~Benyamini,
  L.~Levin, R.~Roth, and O.~Serfaty.
\newblock OFF-set: One-pass factorization of feature sets for online
  recommendation in persistent cold start settings.
\newblock In \emph{Proceedings of the 7th ACM Conference on Recommender
  Systems}, pages 375--378, 2013.

\bibitem[Bykau et~al.(2013)Bykau, Koutrika, and Velegrakis]{BykauVelegrakis13}
S.~Bykau, F.~Koutrika, and Y.~Velegrakis.
\newblock Coping with the persistent cold-start problem.
\newblock In \emph{Personalized Access, Profile Management, and Context
  Awareness in Databases}, 2013.

\bibitem[Chua et~al.(2013)Chua, Oentaryo, and Lim]{chua2013modeling}
F.~C.~T. Chua, R.~J. Oentaryo, and E.-P. Lim.
\newblock Modeling temporal adoptions using dynamic matrix factorization.
\newblock In \emph{IEEE 13th International Conference
  on Data Mining (ICDM)}, pages 91--100, 2013.

\bibitem[Guy et~al.(2009)Guy, Zwerdling, Carmel, Ronen, Uziel, Yogev, and
  Ofek-Koifman]{Guy:2009:PRS:1639714.1639725}
I.~Guy, N.~Zwerdling, D.~Carmel, I.~Ronen, E.~Uziel, S.~Yogev, and
  S.~Ofek-Koifman.
\newblock Personalized recommendation of social software items based on social
  relations.
\newblock In \emph{Proceedings of the Third ACM Conference on Recommender
  Systems}, pages 53--60, 2009.

\bibitem[Hawalah and Fasli(2014)]{Hawalah_2014}
A.~Hawalah and M.~Fasli.
\newblock Utilizing contextual ontological user profiles for personalized
  recommendations.
\newblock \emph{Expert Syst. Appl. (ESWA)}, 41:\penalty0 4777--4797, 2014.

\bibitem[Kluver and Konstan(2014)]{Kluver:2014:ERB:2645710.2645742}
D.~Kluver and J.~A. Konstan.
\newblock Evaluating recommender behavior for new users.
\newblock In \emph{Proceedings of the 8th ACM Conference on Recommender
  Systems}, pages 121--128, 2014.

\bibitem[Knights et~al.(2009)Knights, Mozer, and Nicolov]{KnightsNicolov09}
D.~Knights, M.~Mozer, and N.~Nicolov.
\newblock Detecting topic drift with compound topic models.
\newblock In \emph{Proceedings of the Third International ICWSM Conference},
  pages 242--245, 2009.

\bibitem[Koren(2010)]{koren2010collaborative}
Y.~Koren.
\newblock Collaborative filtering with temporal dynamics.
\newblock \emph{Communications of the ACM}, 53:\penalty0 89--97, 2010.

\bibitem[Rashid et~al.(2002)Rashid, Albert, Cosley, Lam, McNee, Konstan, and
  Riedl]{RashidRiedls02}
A.~M. Rashid, I.~Albert, D.~Cosley, S.~K. Lam, S.~M. McNee, J.~A. Konstan, and
  J.~Riedl.
\newblock Getting to know you: Learning new user preferences in recommender
  systems.
\newblock In \emph{Proceedings of the 7th International Conference on
  Intelligent User Interfaces}, pages 127--134, 2002.

\bibitem[Resnick et~al.(1994)Resnick, Iacovou, Suchak, Bergstrom, and
  Riedl]{ResnickRiedl94}
P.~Resnick, N.~Iacovou, M.~Suchak, P.~Bergstrom, and J.~Riedl.
\newblock Grouplens: An open architecture for collaborative filtering of
  netnews.
\newblock In \emph{Proceedings of the 1994 ACM Conference on Computer Supported
  Cooperative Work}, pages 175--186, 1994.

\bibitem[Saveski and Mantrach(2014)]{SaveskiMantrach14}
M.~Saveski and A.~Mantrach.
\newblock Item cold-start recommendations: Learning local collective
  embeddings.
\newblock In \emph{Proceedings of the 8th ACM Conference on Recommender
  Systems}, pages 89--96, 2014.

\bibitem[Schein et~al.(2002)Schein, Popescul, Ungar, and
  Pennock]{schein2002methods}
A.~I. Schein, A.~Popescul, L.~H. Ungar, and D.~M. Pennock.
\newblock Methods and metrics for cold-start recommendations.
\newblock In \emph{Proceedings of the 25th annual international ACM SIGIR
  conference on Research and development in information retrieval}, pages
  253--260, 2002.

\bibitem[Sedhain et~al.(2014)Sedhain, Sanner, Braziunas, Xie, and
  Christensen]{sedhain2014social}
S.~Sedhain, S.~Sanner, D.~Braziunas, L.~Xie, and J.~Christensen.
\newblock Social collaborative filtering for cold-start recommendations.
\newblock In \emph{Proceedings of the 8th ACM Conference on Recommender
  systems}, pages 345--348, 2014.

\bibitem[Shardanand and Maes(1995)]{ShardanandMaes95}
U.~Shardanand and P.~Maes.
\newblock Social information filtering: Algorithms for automating 'word of
  mouth'.
\newblock In \emph{Proceedings of the SIGCHI Conference on Human Factors in
  Computing Systems}, pages 210--217, 1995.

\bibitem[Shi et~al.(2014)Shi, Karatzoglou, Baltrunas, Larson, and
  Hanjalic]{Shi_cikm_2014}
Y.~Shi, A.~Karatzoglou, L.~Baltrunas, M.~Larson, and A.~Hanjalic.
\newblock Cars2: Learning context-aware representations for context-aware
  recommendations.
\newblock In \emph{Proceeding of CIKM}, pages 291--300, 2014.

\bibitem[Sun et~al.(2012)Sun, Varshney, and Subbian]{sun2012dynamic}
J.~Z. Sun, K.~R. Varshney, and K.~Subbian.
\newblock Dynamic matrix factorization: A state space approach.
\newblock In \emph{IEEE International Conference on Acoustics, Speech and Signal Processing (ICASSP)}, pages 1897--1900, 2012.

\bibitem[Tang et~al.(2014)Tang, Jiang, Li, and
  Li]{Tang:2014:ECB:2645710.2645732}
L.~Tang, Y.~Jiang, L.~Li, and T.~Li.
\newblock Ensemble contextual bandits for personalized recommendation.
\newblock In \emph{Proceedings of the 8th ACM Conference on Recommender
  Systems}, pages 73--80, 2014.

\bibitem[Tavakol and Brefeld(2014)]{Tavakol:2014:FMD:2645710.2645739}
M.~Tavakol and U.~Brefeld.
\newblock Factored mdps for detecting topics of user sessions.
\newblock In \emph{Proceedings of the 8th ACM Conference on Recommender
  Systems}, pages 33--40, 2014.

\end{thebibliography}


\end{document}